\documentclass[%
aps,
prx,
reprint,
superscriptaddress,
nofootinbib,
amsmath,
amssymb
]{revtex4-2}

\usepackage{
    physics,
    graphicx,
    multirow,
    tabularx,
    mathtools, 
    placeins, 
    nicefrac, 
    hyperref, 
    threeparttable, 
    enumitem,
    makecell,
    adjustbox,
    mathrsfs
}

\usepackage{soul}
\usepackage[compat=0.3]{yquant}
\usepackage{quantikz}
\useyquantlanguage{groups}

\usepackage[capitalize]{cleveref}
\Crefname{section}{Sec.}{Secs.}

\hypersetup{
    colorlinks,
    linkcolor={blue!50!black},
    citecolor={blue!50!black},
    urlcolor={blue!80!black}
}

\usepackage[caption=false]{subfig} 
\begin{document}
\title{
Nanostructure modelling with early fault tolerant quantum computers}

\newcommand{\qmaddress}{\affiliation{Quantum Motion, 9 Sterling Way, London N7 9HJ, United Kingdom}}
\newcommand{\oxddress}{\affiliation{Department of Materials, University of Oxford, Parks Road, Oxford OX1 3PH, United Kingdom}}
\newcommand{\mathinst}{\affiliation{Mathematical Institute, University of Oxford, Woodstock Road, Oxford OX2 6GG, United Kingdom}}

\author{Zhu Sun}
\email[zhu.sun@exeter.ox.ac.uk]{}
\oxddress
\mathinst
\qmaddress

\author{Christian Binder}
\oxddress
\qmaddress

\author{B\'alint Koczor}
\mathinst
\qmaddress

\author{Simon Benjamin}
\oxddress
\qmaddress

\begin{abstract}

Semiconductor nanostructures are central to many developing technologies. Notably, double quantum dots are especially important for semiconductor spin-qubit architectures, quantum sensing applications, and quantum-dot solar cells. Accurate modelling is highly desirable but conventional methods can struggle when dynamics involve more than two interacting electrons. In this work, we present a quantum simulation framework capable of addressing multi-electron double quantum dots. We adopt an efficiently scaling 1$^\text{st}$ quantised representation of the system and develop algorithms based on both Trotterisation and Qubitisation. Incorporating insights from classical simulations enables us to produce resource estimates that are more realistic than those obtained from theoretical error bounds. Using a standard surface code model with physical noise at $10^{-3}$, our results indicate that the ground-state energy of four electrons in a double quantum dot can be estimated in approximately 22 hours using 226k physical qubits, or an eight-electron system in 3.3 days with 314k qubits (with runtimes falling dramatically when more qubits are available). We anticipate that incorporating recent advances in surface code architectures may reduce these costs significantly further. Our results suggest that early fault-tolerant quantum computers may become valuable tools for designing mature-era quantum technologies.

\end{abstract}

\maketitle

\section{Introduction}

Semiconductor nanostructures, such as quantum wells, nanowires, and quantum dots, underpin a wide range of modern technologies, ranging from lasers and light-emitting diodes to high-efficiency solar cells and nanoscale sensors~\cite{Kvon2023SemiconductorNanostructures,GarciaDeArquer2021SemiconductorQDs}. Accurate modelling of the electronic structure of these systems is therefore of both fundamental and practical importance, as it enables the prediction of optical spectra, transport properties, and device performance directly from the underlying confinement potential and material parameters. Among these platforms, quantum dots play a particularly prominent role. For example, arrays of silicon quantum dots have been explored as active elements in high-efficiency solar cells, where quantum confinement facilitates band-gap engineering for tandem architectures~\cite{conibeer_solar}. They have also been proposed and employed as sensitive quantum sensors, for example in nanoscale magnetic-field sensing using driven double quantum dots (DQD) \cite{giavaras_sensing}. Finally, silicon quantum dots are also a leading candidate for qubit implementations, with electrons confined by gate-defined potentials serving as the fundamental quantum degrees of freedom~\cite{loss1998quantum,burkard_rmp}. 

All of these applications ultimately rely on electrons in silicon that are confined by electrostatic gate potentials. Accurately predicting their wave functions and energy spectra is therefore essential both for understanding existing devices and for guiding new designs. For single-electron dots, effective-mass or envelope-function methods enable numerically exact simulations for realistic device geometries on classical computers within seconds~\cite{binder_envelope}. However, adding more electrons leads to a rapid increase in computational cost. Even for just two interacting electrons in realistic three-dimensional gate-defined potentials, state-of-the-art tight-binding and real-space configuration-interaction calculations already demand substantial computational resources, ranging from minutes to hours on a single CPU~\cite{binder_envelope}. For many of the applications mentioned above, the relevant operating regime involves more than two electrons per dot, making the development of methods capable of accessing the few- to many-electron regime a central challenge.

Within this broader context, semiconductor spin qubits hosted in quantum dots represent a particularly promising route toward large-scale quantum computation~\cite{loss1998quantum,burkard_rmp}. Quantum-dot spin qubits are compatible with advanced semiconductor manufacturing and have demonstrated high-fidelity single- and two-qubit gates in silicon and related materials \cite{PhysRevApplied.21.014044,doi:10.1126/sciadv.abn5130,noiri2022fast}. While the original Loss–DiVincenzo proposal focused on single-electron spins, modern qubit encodings frequently employ multi-electron quantum dots, for example, three-electron exchange-only and resonant-exchange qubits in triple dots~\cite{russ_three_electron,medford_rx}, as well as three-electron states in double dots~\cite{shi_three_electron}. Semiconductor qubits already demonstrate sub-microsecond gates~\cite{Petta2022} and rapid shuttling~\cite{LievenShuttle} with fidelities surpassing the threshold for fault tolerant QEC; however the governing physics permits even faster operations and reaching the ultimate performance possible in mature systems may require deep and comprehensive modelling of many-electron structures. A plausible scenario is that approximate classical methods are used as a rapid design tool, but validation with more exact quantum simulation is employed prior to committing to a chip design tapeout and manufacture. In such a use case, runtimes of hours or days for the quantum algorithm  could be acceptable.

In this work, we focus on a gate-defined DQD potential in silicon as a representative example of a quantum-dot device. We develop a quantum algorithm for simulating multiple electrons in a DQD potential by adopting a grid-based representation of the DQD system. As such, our approach is formulated in the first-quantised framework, in contrast to the second-quantised methods that are more commonly explored in the quantum simulation literature. In general, the resource costs for first-quantised methods are excepted to scale more efficiently with the number of simulated particles \cite{PRXQuantum.2.040332,huang2025fullqubit}, although the baseline cost for a given small simulated system may be higher than in second quantisation. We are therefore pleased with the fairly modest resource costs for our small 4 and 8 electron systems, and we expect that the approach will remain competitive and efficient for larger target systems.

Grid-based classical simulation of quantum systems have been extensively explored for decades \cite{PhysRevE.73.036708,cerjan2013numerical,light2000discrete}. Extending these approaches to quantum simulation is conceptually straightforward, and corresponding grid-based quantum simulation frameworks have been developed \cite{doi:10.1073/pnas.0808245105,PhysRevLett.125.260511,PhysRevResearch.4.033121,doi:10.1126/sciadv.abo7484}. A grid-based classical method for simulating quantum dots was explored in ref.~\cite{rm43-nh9b}, though the analysis there is limited to two-electron systems. In contrast, our approach can efficiently extract information from multi-electron systems in both static (e.g., exchange coupling) and dynamic (e.g., shuttling) scenarios. Within the grid-based framework, we investigate two techniques for estimating the ground-state energy: Trotter-based quantum phase estimation (QPE) and qubitisation. For both methods, we provide resource estimates with variants ranging from qubit-frugal to runtime-minimising.

We note that, in the case of the Trotter method, our algorithm closely mirrors the approach taken in certain of state-of-the-art classical methods~\cite{binder_envelope,rm43-nh9b}. In particular, the Trotter step size $\delta t$ in our method can be directly compared with the time-step granularity used in classical simulations. Whereas analytic error bounds on $\delta t$ for quantum algorithms can be stated, we believe they are overly conservative since the analogous step size $\delta t$ used successfully in classical simulations is orders
of magnitude larger. We argue these larger step sizes $\delta t$ should also be adequate for our quantum algorithm. Moreover, we contend that similar choices of $\delta t$ remain appropriate as the quantum simulation scales from the commonly studied two-electron case ($n_p = 2$) to larger systems such as $n_p = 8$. We provide explicit expressions for the quantum resources required for general $n_p$ and evaluate these resource estimates for $n_p = 4$ and $n_p = 8$.

Assuming quantum hardware consisting of a square grid of qubits with nearest-neighbour connectivity, a physical error rate of $10^{-3}$, and a code cycle time of $1~\mu s$, we estimate the time and qubit resources required to obtain a ground-state energy. In our most qubit-frugal approach, the $n_p = 4$ case requires approximately 22 hours and $226\text{k}$ physical qubits. We also present several spacetime trade-offs in which increased qubit counts yield reduced runtimes. Furthermore, a problem with $n_p = 8$ electrons can be solved in roughly in 78 hours given access to 314k qubits.


The remainder of this manuscript is structured as follows. In \cref{sec:background}, we review the relevant background on grid-based simulation and phase oracles. We then present quantum algorithms for estimating the ground state energy using Trotter-based QPE (\cref{subsec:trotter}) and qubitisation (\cref{subsec:qubit}). Potential applications of these algorithms are discussed in \cref{sec:applications}, and concluding remarks are given in \cref{sec:conclusion}. Additional implementation details are provided in the Appendices.

\section{Background} \label{sec:background}

\subsection{Grid-based Simulation}

The key idea in first-quantised grid-based Hamiltonian simulation is to discretise the $D$-dimensional space into uniformly spaced grid points. This discretization imposes a finite spatial resolution on the quantum states, which, under the quantum Fourier transform (QFT), is equivalent to imposing a finite momentum cutoff in $k$-space. Consequently, the basis functions in real-space are not the Dirac delta functions but rather ``smeared'' versions of it, taking the form of Dirichlet kernels, which closely resemble $\text{sinc}(x)$ functions \cite{rm43-nh9b}. Each grid point therefore occupies a finite ``volume'' in $D$-dimensions.

The method we use to implement the dynamics is the split-operator QFT (SO-QFT) Hamiltonian simulation \cite{doi:10.1073/pnas.0808245105,PhysRevLett.125.260511,PhysRevResearch.4.033121,doi:10.1126/sciadv.abo7484}, which is based on the Suzuki–Trotter formula (also known as Trotterisation). In Trotterisation, the time-evolution operator over a total simulation time $t$ is decomposed into $N$ small time steps $\delta t=t/N$. For a Hamiltonian $H$, this yields
\begin{equation}
    e^{-iHt}=(e^{-iH\delta t})^N
\end{equation}

Applying, for instance, the second order Suzuki-Trotter formula for $H=H_{kin}+H_{pot}$, where $H$ decomposed into kinetic and potential components, we obtain 
\begin{equation} \label{eqn:2nd}
    e^{-iH\delta t}=e^{-iH_{kin}\delta t/2}e^{-iH_{pot}\delta t}e^{-iH_{kin}\delta t/2} + O(\delta t^3)
\end{equation}

The key observation that simplifies the simulation for the grid-based representation is that $H_{kin}$ is diagonal in $k$-space, while $H_{pot}$ is approximately diagonal in real space (due to the finite spatial resolution) -- and diagonal unitaries can be implemented as quantum circuits using standard techniques. Furthermore, the QFT, which allows us to switch between the two representations, admits efficient implementation as being one of the most well studied quantum algorithmic primitives. For an extended introduction to the theoretical framework of first-quantized grid-based Hamiltonian simulation, we refer the reader to ref.~\cite{doi:10.1126/sciadv.abo7484}.

\subsection{The Phase Oracle and its variants} \label{sec:oracle}

In this section, we introduce the key tool used to apply bespoke phase according to the potential terms is $H$, namely the phase oracle. Mathematically, a phase oracle implements the mapping $\ket{x} \mapsto e^{i f(x)} \ket{x}$ for some function $f(x)$ associated with the computational basis state $\ket{x}$. Ref.~\cite{m32k-7nq2} developed an efficient construction of phase oracles for functions $f(x)$ that can be approximated using a piecewise polynomial function. For each segment of the piecewise function, the circuit first computes a flag indicating whether the value stored in the input register lies within that segment, then applies $R_Z$ rotations to implement the corresponding phase factor, and finally uncomputes the flag. Using techniques such as fan-out, if $f(x)$ is approximated by $S$ polynomial segments, their circuit can apply unique phase functions to all $S$ pieces in parallel via single qubit rotations at the cost of an increased qubit count by a factor of $O(S)$. This $S$-fold parallelization was employed in that work because the authors sought to minimize circuit depth. More generally, however, the architecture of the phase oracle circuit naturally supports adjustable degrees of parallelization---from full $S$-fold parallelization to a completely serialized implementation that processes the $S$ pieces sequentially.

In addition to the phase oracle circuit presented in figure 1 of ref.~\cite{m32k-7nq2}, which we refer to as the ``ROT1'' phase oracle, the authors also considered a quantum read-only memory (QROM)--based variant. The QROM phase oracle uses QROM \cite{PhysRevX.8.041015} to load the parameters appearing in the polynomial interpolation of $f(x)$ \cite{PRXQuantum.1.020312}, then evaluates the function via arithmetic operations, and finally applies the phase by performing an addition into a phase gradient state. As in the ROT1 case, the QROM phase oracle was originally constructed with $S$-fold parallelisation, but can also be fully serialized. The relative performance of ROT1 and QROM phase oracles, in terms of circuit depth and qubit count, is highly application-dependent and typically requires case-by-case analysis.

Ref.~\cite{m32k-7nq2} also pointed out that, by adding only a few Clifford gates, the ROT1 phase oracle circuit can be modified to operate as an amplitude oracle, which implements the mapping $\ket{0}\ket{x}\mapsto(\cos{f(x)}\ket{0}{+}i \sin{f(x)}\ket{1})\ket{x}$ for some smooth function $f$ that can be piecewise approximated. We will demonstrate how to use this amplitude oracle for block encoding in \cref{subsec:qubit}.

\section{Methods} 

Consider a $D$ dimensional simulation domain with volume $L^D$ containing $n_p$ electrons. The Hamiltonian of the system can be written as

\begin{equation}\label{eqn:ham}
\begin{split}
H=
& \underbrace{\sum_{i=1}^{n_p}\frac{\hbar^2}{2m} \mathbf{k}_i^2}_{\text{kinetic}}
+ \underbrace{\sum_{i=1}^{n_p}U_{DQD}(\mathbf{r}_i)+g(\mathbf{r}_i)\mu_B \mathbf{B\cdot S}_i\ }_{\text{DQD potential}}\\
& + \underbrace{\sum_{i,j=1;i\neq j}^{n_p} \frac{e^2}{4\pi \epsilon_0\epsilon_r |\mathbf{r}_i-\mathbf{r}_j|}}_{\text{Coulomb potential}}
\end{split}
\end{equation}
where we partition $H$ into kinetic part $K$, DQD potential $U$ and Coulomb potential $V$. Here $\mathbf{r}_i$ represents the position of the $i$-th electron within the $D$-dimensional simulation domain, $g(\mathbf{r}_i)$ is the position-dependent g-factor of electron $i$, while $\mathbf{S}_i$ represents the respective electron spin operator. The relative permittivity $\epsilon_r$ is $11.9$ for the dielectric conditions of silicon.

Silicon quantum dots are three-dimensional regions that are relatively narrow in the vertical direction, resulting in a characteristic "pizza-box" shape. Consequently, it is efficient to employ an effective two-dimensional envelope-function theory~\cite{binder_envelope}, replacing the full three-dimensional problem with a quasi-two-dimensional description in the lateral $(x,y)$ plane. In this approximation, the motion perpendicular to the interface (the $z$-direction) is incorporated into an effective two-dimensional potential. The approach also employs an effective in-plane Coulomb interaction, so setting $D = 2$ provides quantitatively accurate results for realistic devices. Therefore, we can represent a single particle state within the grid-based method as 
\begin{equation}\label{eqn:state}
\sum_{r_x,r_y=-\rho}^{\rho-1} \sum_{s=0}^1 \Psi(r_x,r_y,s) \ket{r_x}\ket{r_y}\ket{s}
\end{equation}
where we define $\rho = 2^{n_q-1}$. Therefore, a three-dimensional DQD system containing $n_p$ particles can be represented by a $n_p(2n_q+1)$-qubit state.  Note that if one wished to model nanostructures where the $z$-direction is not small compared to the $x$-$y$ dimensions, then one could simply assign an $\ket{r_z}$ register in \cref{eqn:state} -- the logical qubit count associated with the electron state representation then increases by a fixed fraction (e.g. $50\%$) regardless of the number of modelled electrons.

In the following sections, we elaborate on the methods we used, namely Trotterisation and Qubitisation, for estimating the ground state energy of this Hamiltonian. For both methods, we first describe the implementation details and then analyse the cost for the case $n_p = 4$ and $n_q = 8$. At the fault-tolerant level, our core metrics are measurement depth, Toffoli count and additional ancilla required. We assume that each T-gate, Toffoli gate and measurement contributes a measurement depth of 1, measured in the unit of $d$ code cycles. While some literature reports results in terms of T-count, following ref.~\cite{Gidney2019efficientmagicstate}, we use the conversion $\text{T-count}/2 = \text{Toffoli count}$. The ``extra ancilla'' metric accounts for auxiliary qubits required by subroutines such as adders, and is intended to capture the maximum qubit usage (i.e., the widest part of the circuit) throughout the algorithm. We minimize this by reusing ancilla where possible, without compromising efficiency. Further technical details, including analytical expressions for resource estimation, are given in \cref{appx:Trotter} and \cref{appx:Qubit}.

\subsection{Trotterisation} \label{subsec:trotter}
In this work, we use second order Trotterisation. According to \cref{eqn:2nd}, the corresponding SO-QFT time propagator is
\begin{equation} \label{eqn:SOQFT}
\begin{aligned}
e^{-iH\delta t}&\approx e^{-iK\delta t/2}e^{-i(U+V)\delta t}e^{-iK\delta t/2},\\
\text{with } e^{-iK\delta t/2} &= QFT^\dagger e^{-iD_{K}\delta t/2}QFT
\end{aligned}
\end{equation}
where $D_K$ is $K$ in the $k$-space representation and it is diagonal. The potentials $U$ and $V$ are approximately diagonal in real space.

The strategy is to apply phase corresponding to the $K$, $U$ and $V$ separately, which we refer to as stage A, B and C. A schematic circuit diagram for one Trotter step is shown in \cref{fig:trotter_step}. As we detail below, the phase related to $e^{-iD_{kin}\delta t/2}$ can be applied relatively cost efficiently when given $\mathbf{k}_i$. In contrast, the functions of the form $f(\mathbf{r}_i)$ and $f(\mathbf{r}_i,\mathbf{r}_j)$ require a more involved construction for given $\mathbf{r}_i$.

\makeatletter

\DeclareRobustCommand\rvdots{%
\vbox{%
\baselineskip4\p@\lineskiplimit\z@%
\kern-\p@%
\hbox{.}\hbox{.}\hbox{.}%
}%
}

\begin{figure*}
    \centering

\begin{tikzpicture}
\begin{yquant*}

nobit t;
hspace {2mm} t;
text {Stage A} t;
hspace {2mm} t;
text {Stage B} t;
hspace {27mm} t;
text {Stage C} t;

init {Particle 1} (a[0-3]);
discard a[2];
text {{$\rvdots$}} a[2];

init {Particle 2} (b[0-3]);
discard b[2];
text {{$\rvdots$}} b[2];

init {Particle 3} (c[0-3]);
discard c[2];
text {{$\rvdots$}} c[2];

init {Particle 4} (d[0-3]);
discard d[2];
text {{$\rvdots$}} d[2];

box {KIN} (a[0-3]);
box {KIN} (b[0-3]);
box {KIN} (c[0-3]);
box {KIN} (d[0-3]);

[red, thick] barrier (a,b,c,d);
box {POT1} (a[0-3]);
box {POT1} (b[0-3]);
box {POT1} (c[0-3]);
box {POT1} (d[0-3]);

[red, thick] barrier (a,b,c,d);
box {POT2} (a[0-3],b[0-3]);
box {POT2} (c[0-3],d[0-3]);
box {POT2} (a[0-3],c[0-3]);
box {POT2} (b[0-3],d[0-3]);
box {POT2} (a[0-3],d[0-3]);
box {POT2} (b[0-3],c[0-3]);


hspace {11.7mm} a[2];
text {{$\rvdots$}} a[2];
text {{$\rvdots$}} b[2];
text {{$\rvdots$}} c[2];
hspace {11.7mm} d[2];
text {{$\rvdots$}} d[2];

\end{yquant*}
\end{tikzpicture}
\caption{A schematic circuit diagram for one Trotter step for $n_p=4$ particles. The KIN, POT1 and POT2 operations correspond to the $K$, $U$ and $V$ terms acting on one- and two-particle registers, respectively. Note that the POT2 operations in stage C can be executed in 3 layers.}
\label{fig:trotter_step}
\end{figure*}
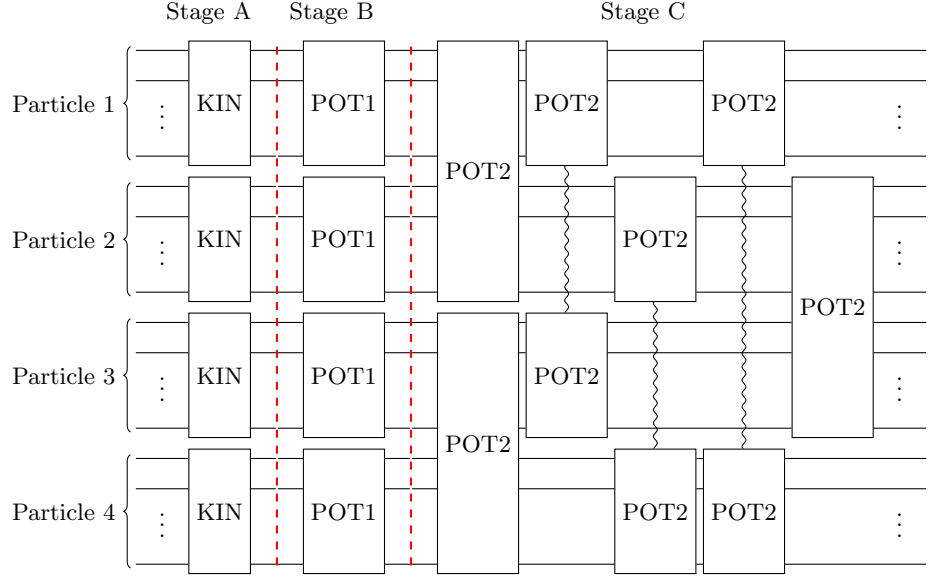

The QFT and its inverse are implemented by preparing a reusable phase gradient state and applying arithmetic addition circuits to it. Note that when the Hamiltonian is partitioned into two terms, as in our case, the 2nd order Trotter circuit is nearly identical to the 1st order Trotter circuit, while still offering improved error scaling \cite{PhysRevLett.128.210501}. Consequently, although our simulations employ 2nd Trotterisation, we use the resource cost of the 1st Trotter circuit in our resource estimates.

Implementing the phase shifts corresponding to $K$, or more precisely $D_K$, reduces to applying a phase of the form $\exp(i\alpha k^2)$, where $\alpha = -\frac{\delta t \hbar^2}{2m}$ and $k$ is stored in a quantum register. The computation of $k^2$ can be performed using binary arithmetic. As an illustration, consider the following two-bit example:
\begin{equation} \label{eqn:square}
\begin{aligned}
(10)_2^2&=(1\times2^1+0\times2^0)^2 \\
&= (1\times2^1)^2+(0\times2^0)^2+2(1\times2^1)(0\times2^0)
\end{aligned}
\end{equation}

The pattern of this expansion suggests the phase can be implemented using single qubit $Z$-rotation gates ($R_Z$) and controlled-$Z$-rotation gates ($CR_Z$). To realize the square terms in \cref{eqn:square}, we apply $R_Z(2^0\alpha)$ and $R_Z(2^2\alpha)$. To realize the cross terms, we apply $CR_Z(2^2\alpha)$. The generalization to an $n_q$-qubit register is straightforward: one applies $n_q$ single-qubit $R_Z$ gates and $n_q(n_q-1)/2$ two-qubit $CR_Z$ gates. Note that the $CR_Z$ gate is symmetric with respect to its control and target qubits; thus, either qubit involved can serve as the control or target.

Next, the DQD potential can be approximated as a two-dimensional potential $U_{xy}$, in accordance with the envelope function theory. We further approximate this potential using a piecewise-linear representation, 
\begin{equation}
U_{xy}^{(i)}(x,y) = a_ix+b_iy+c_i, \text{  for } i\in \{1,2,...,S_1\}    
\end{equation}
this allows a 2D phase oracle to be applied. We find $S_1=42$ is sufficient to approximate the function with the desired accuracy (see \cref{appx:Trotter}), and we employ a fully serialized QROM phase oracle for this purpose. Note that the value of $S_1$ does not explicitly depend on the choice of $n_p$ and $n_q$. 

Finally, applying phase corresponding to the Coulomb potential requires first computing the squared distance $r^2$ between two particles. This is achieved through quantum arithmetic: we first compute the coordinate differences $\Delta x$ and $\Delta y$ via subtraction, then square and sum them to obtain $r^2$. A fully serialized ROT1 phase oracle with $S_2=8$ pieces is then applied to the register storing $r^2$ to implement the phase factor $\exp(if(x))$, where $f(x) \propto 1 / \sqrt{x}$. This procedure is repeated for every particle pair, that is, for $n_p(n_p - 1)/2$ pairs in a system of $n_p$ particles.

The procedures described above implement a single Trotter step. For QPE, a controlled version of the Trotter-step operator is required. Fortunately, the additional cost of introducing this control is marginal, and the corresponding method is detailed in \cref{appx:Trotter}.

We now evaluate the resource required for the case $n_p = 4$ and $n_q = 8$. A breakdown of the cost for the controlled-Trotter-step operator is provided in \cref{tab:breakdown}.

\begin{table}[h]
    \centering
    \begin{tabular}{c|c|c|c}
                & Depth & Ancilla & Toffoli \\
        \hline
        Stage A & 349   & 128     & 3569 \\
        \hline
        Stage B & 1172  & 198     & 8484 \\
        \hline
        Stage C & 1863  & 144     & 17011  \\
        \hline 
        Overall & 3384  & 198     & 29064
    \end{tabular}
    \caption{Cost breakdown for implementing a controlled-Trotter step operator for the case $n_p = 4$ and $n_q = 8$. The cost metrics are measurement depth, extra ancilla count, and Toffoli count.}
    \label{tab:breakdown}
\end{table}

To provide logical resource estimates, we begin by outlining several key assumptions. Firstly, we assume a rotated surface code with lattice surgery as the quantum error-correcting (QEC) scheme, implemented on a square grid of physical qubits with nearest-neighbour connectivity. 

Next, in what follows, we first assume a beat-limited computation, meaning the computation speed is constrained by the rate of magic state consumption, which is why we previously counted the measurement depth. In the beat-limited regime, the execution time of Clifford gates becomes relevant, in contrast to the tick-limited model used, for example, in ref.~\cite{PRXQuantum.2.030305}, where the computation rate is constrained by magic state production. However, since the measurement depth accounts only for the time associated with magic state operations, we assume a 1:1 ratio between magic depth and Clifford depth, i.e., on average, each layer of magic state operations is followed by a layer of Clifford gates. Therefore, the total depth for one Trotter step is 6768. Although  state-of-the-art lattice surgery techniques require approximately $2d$ cycles for a $CNOT$ gate~\cite{horsman2012surface}, $3d$ cycles for the Hadamard ($H$) gate~\cite{litinski2019game}, and $d$ cycles for the phase ($S$) gate~\cite{hirai2026boldsymbol2dtimesdtimes}, ref.~\cite{hirano2025locality} proposed that locality-aware Pauli-based computation (LAPBC) can be applied during compilation. In brief, LAPBC propagates and removes all single-qubit Clifford gates, i.e., the $H$ and $S$ gates are eliminated, while the CNOT gates are retained. This approach preserves both the locality and parallelism of the original circuit, in contrast to conventional Pauli-based computation schemes (see, e.g., ref.~\cite{litinski2019game}). Overall, we regard the assumed magic-to-Clifford depth ratio as a reasonable estimate under this framework.

For the magic state factory, we assume the 8T-to-CCZ distillation factory described in \cite{gidney2025factor}. By leveraging magic state cultivation \cite{gidney2024magic}, this factory operates on 12 logical qubits and produces one CCZ state in approximately $5d$ code cycles. To simplify the analysis, we assume a steady rate of magic state consumption. Thus, the average number of factories required to sustain a beat-limited computation is  $29064/6768\times5\approx21.5$, corresponding to an average of 257 logical qubits. We also allocate an equal number of logical qubits (i.e., another 257) to account for routing overhead. 

\begin{figure*}
    \centering
    \includegraphics[width=.5\textwidth, trim=1.3cm 2.1cm 1.3cm 1.2cm, clip]{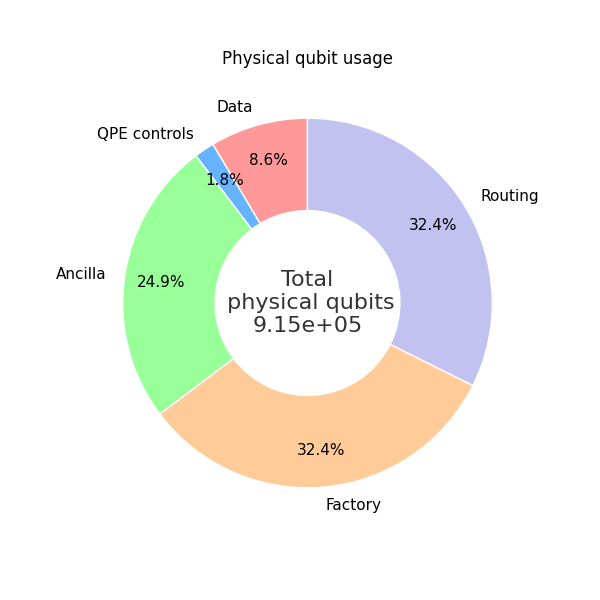}\hfill
    \includegraphics[width=.5\textwidth, trim=1.3cm 2.1cm 1.3cm 1.2cm, clip]{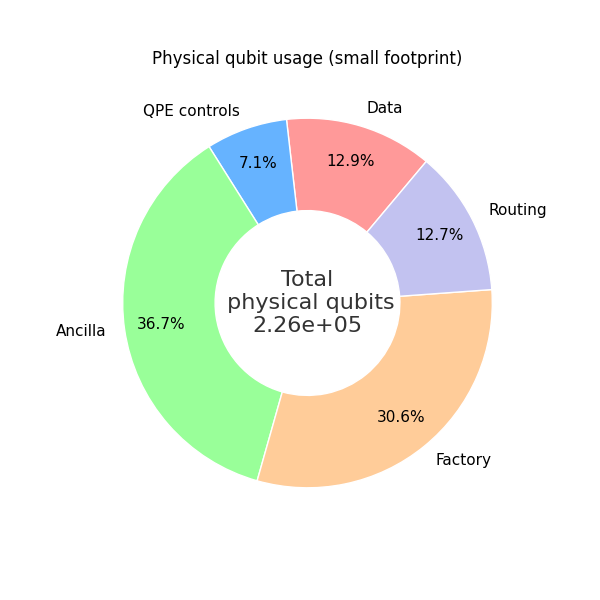}
    \caption{The logical qubit usage breakdown for $n_p=4$ particles, illustrated as pie charts. Left: Pie chart for the highly parallelized, beat-limited computation. Right: Pie chart for the tick-limited computation employing only five magic state factories and yoked surface codes for data qubit storage, referred to as the small-footprint scheme.}
    \label{fig:pie}
\end{figure*}

The total logical qubit count further includes 68 data qubits representing the system’s state and 14 control qubits required for the QPE algorithm, and these all sum up to 794 logical qubits. A detailed breakdown of logical qubit usage for the case $n_p = 4$ and $n_q = 8$ is shown in the pie chart in \cref{fig:pie}. Note that the ancilla metric represents the maximum number of auxiliary qubits required throughout the algorithm. This number is not always saturated, so inactive ancilla qubits may, for instance, be repurposed to assist with magic state production.

To convert these numbers to physical resources, we assume the physical qubits have an error rate $0.001$ and $1\mu s$ code cycle time. Although the Trotter error bound suggests that up to $4\times10^{11}$ Trotter steps are required to achieve the desired accuracy, classical simulations for the $n_p = 2$ case \cite{binder_envelope} indicate that QPE can estimate the ground-state energy with only $\sim10^5$ steps (see \cref{appx:Trotter} for details). For the code distance $d$, we show in \cref{appx:Trotter} that $d=23$ is sufficient for this computation. With these numbers, we can estimate the physical qubit count for the case $n_p = 4$ and $n_q = 8$, which is $\sim915$K. The QPE algorithm has runtime 4.32 hours but each shot has a logical-error-free rate $\sim88.4\%$, thus, the expected runtime is 4.89 hours.

\begin{figure*}
    \centering
    \includegraphics[width=0.85\linewidth]{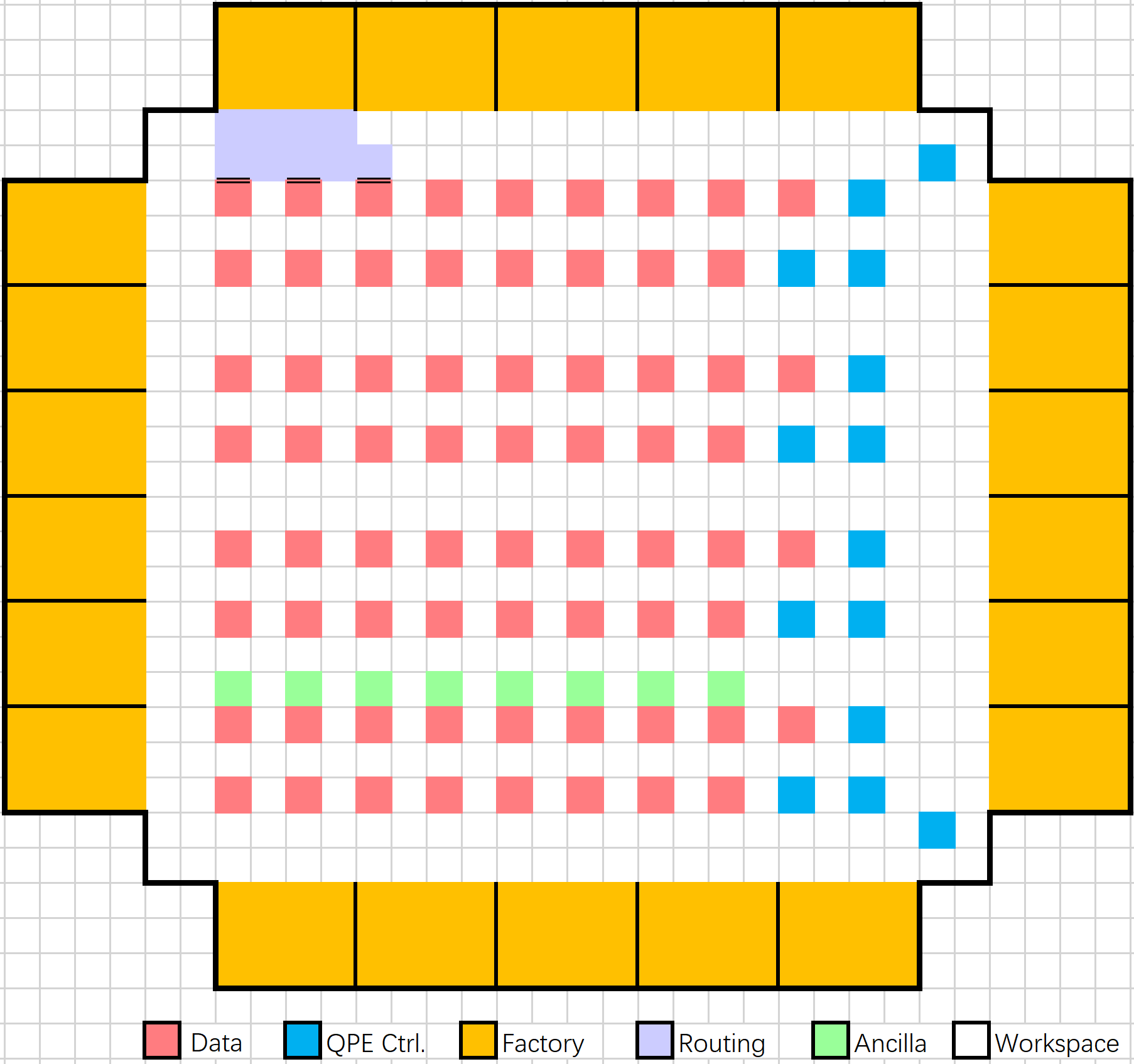}
    \caption{A possible layout of logical qubits for the case $n_p=4$ and $n_q=8$ on rotated surface code, assuming the high parallelism beat-limited computation. The red patches denote data qubits storing the quantum state of the system, while the blue patches represent the control qubits for QPE. Each 8T-to-CCZ factory comprises 12 orange patches, separated by thin black lines. The purple region at the top illustrates an example of routing for teleporting a CCZ state; the double lines indicate measurements at the boundaries. The green patches at the bottom are active ancilla, showing an example of implementing 8 single-qubit rotations in parallel. The white patches represent workspace qubits that can be used for routing or additional ancilla as needed. The entire computing region is enclosed by the thick black line and consists of 792 squares.
    \label{fig:layout}}
\end{figure*}

A possible layout of logical qubits for the $n_p = 4$, $n_q = 8$ case using the rotated surface code is shown in \cref{fig:layout}, assuming beat-limited computation. The region enclosed by the thick black boundary consists of 792 squares, each representing a surface code patch.

The qubit count can be reduced by slowing down the rate of magic-state consumption, as this lowers both factory requirements and routing overhead. If the computation is slowed by a factor of 2 through serializing, the QPE runtime increases to 8.65 hours, while the logical qubit count decreases to 437, corresponding to approximately $503$K physical qubits at a physical error rate of 0.001. The probability of a logical-error-free simulation is 87.3\%, and thus the expected runtime is 9.91 hours. We note that the decrease in the logical-error-free rate implies that introducing a slowdown factor increases the spacetime volume, since the number of data qubits required are unaffected by this slowdown.

The qubit count can be further compressed by adopting a ``small footprint” scheme, as proposed in, e.g., refs.~\cite{PRXQuantum.2.030305,gidney2025factor}, in which the number of distillation factories is only sufficient to produce one magic state every $d$ code cycles. Under this scheme, because only one magic state is consumed every $d$ code cycles, the routing overhead can be reduced and idle qubits can be packed more densely. We store idle qubits using yoked surface code patches \cite{gidney2025yoked}. Following figure~6 of ref.~\cite{gidney2025factor}, the required number of physical qubits per logical qubit can be reduced to approximately one third of that required by standard surface code patches while still achieving the target logical error rate. We allocate an equal number of physical qubits for routing. With these modifications, QPE requires only 226k physical qubits, placing it within the regime of early fault-tolerant quantum devices \cite{PRXQuantum.5.020101}, although the expected runtime increases to 21.78 hours. The corresponding logical-qubit usage pie chart is shown in \cref{fig:pie}. In this small footprint scheme, one could employ a layout similar to that shown in figure~7 of Ref.~\cite{gidney2025factor}. A summary of the resource estimates is provided in \cref{tab:results}.

As the number of particles $n_p$ increases, there is a linear increase in the implementation cost of the Coulomb interaction part of each Trotter cycle, as is evident from \cref{fig:trotter_step}. One might also anticipate that, as more electrons are added to a confined potential, the duration simulated by a single Trotter step \(\delta t\) would need to be reduced -- thereby increasing the total number of Trotter steps required to model a given process. However, within the parameter regime considered here, the many-body ground state remains close to a Slater determinant constructed from a small set of single-particle orbitals; in other words, the system does not develop strong correlations \cite{InPrep}. Consequently, the occupied orbitals have a similar extent in momentum space as in the one- and two-electron cases, and it is precisely this \(k\)-space extent that determines the Trotter error and hence the allowable Trotter step size. We therefore assume that for $n_p=8$ particles, the same number of Trotter steps yields comparable accuracy to the $n_p=4$ case. Under this assumption, the expected runtime for $n_p=8$ particles is 8.96 hours, requiring approximately 2 million qubits. As before, one can slow the computation down by e.g., a factor of 2, in which case the expected runtime becomes 18.00 hours, requiring approximately 1 million qubits. Note that the code distance $d$ increases to 25 in this case. See \cref{tab:results} for more details. The relatively modest scaling of resource costs from $n_p=4$ to $n_p=8$ is consistent with the generally-held view that the first-quantised methods offer superior asymptotic scaling, albeit their constant factors have been expected to be substantial. When scaling to higher $n_p$ we would expect the resource cost scaling to remain modest while the assumptions made here remain valid (including insensitivity of $\delta t$ to $n_p$).

\begin{table*}
    \centering
    \begin{tabular}{c|c|c|c|c|c}
    & Logical q. & Physical q. & Runtime (h) & Error-free rate & Exp. runtime (h) \\
        \hline
         $n_p=4$, beat limit& 794 & $9.15\cdot10^5$ & 4.32 & 0.884 & 4.89 \\
         $n_p=4$, x2 slowdown& 437 & $5.03\cdot10^5$ & 8.65 & 0.873 &9.91\\
         $n_p=4$, small footprint& 239& $2.26\cdot10^5$& 18.57&0.852&21.78\\
         $n_p=8$, beat limit& 1536 & $2.08\cdot10^6$ & 8.55 & 0.954 & 8.96\\
         $n_p=8$, x2 slowdown& 843 & $1.14\cdot10^6$ & 17.09& 0.949& 18.00\\
         $n_p=8$, small footprint&325& $3.14\cdot10^5$& 71.87& 0.919&78.17\\
    \end{tabular}
    \caption{Summary of the resources required for the different instantiations. Runtimes are given in hours. Note that, in the small footprint scheme, data qubits are encoded using yoked surface codes, so the physical qubit cost is not uniform across logical qubits.}
    \label{tab:results}
\end{table*}

Finally, we emphasize that the implementation of this algorithm is guided by the need to balance qubit count reduction against runtime optimization. Nevertheless, the algorithm can be further accelerated at the cost of additional qubits, for example by using optimal-depth adders or by employing $S_i$-fold parallelization of the phase oracles. Conversely, the qubit count can be reduced further by, for example, decreasing the number of magic state factories and allocating less routing space, albeit at the cost of increased execution time.

\subsection{Qubitisation} \label{subsec:qubit}

We first construct the block encoding of $H$. Consider rewriting \cref{eqn:ham} as 
\begin{equation}\label{eqn:ham2}
H=\sum_{i=1}^{n_p}K_x^{(i)}+\sum_{i=1}^{n_p}K_y^{(i)}+\sum_{i=1}^{n_p}U^{(i)}+\sum_{i,j=1;i\neq j}^{n_p} V^{(ij)}
\end{equation}

Our strategy for block encoding is to block encode the $n_p^2/2+5n_p/2$ terms in \cref{eqn:ham2} separately and then use linear combination of unitary (LCU) \cite{10.5555/2481569.2481570} to combine them.

Since the system's state is encoded in real space, to block encode the kinetic part, we first apply QFT to get the momentum. We then use arithmetic to compute and store $k^2$. Next, we apply bitwise $CR_Y$ gates to rotate an ancilla, so that the amplitude now contains information corresponding to $k^2$. We also need to uncompute $k^2$ and the QFT.

We then use a fully serialized amplitude oracle to block encode the potentials. Recall that for the amplitude oracle from \cref{sec:oracle}, the output of the amplitude oracle for a 2D function is in the form $(\cos{f(x,y)}\ket{0}{+}i \sin{f(x,y)}\ket{1})\ket{x}\ket{y}$. Therefore, to encode a DQD potential $U_{xy}$, we set $f=\arccos(\tilde{U}_{xy})$, where $\tilde{U}_{xy}$ is the normalized version of $U_{xy}$. Block encoding the Coulomb follows a similar procedure.

We use the PREPARE and SELECT method \cite{low2018hamiltonian} for LCU. For a Hamiltonian in the form 
\begin{equation}
    H=\sum_j c_j H_j
\end{equation}
where $c_j$ are positive coefficients and $H_j$ are unitary operators we can then define operators
\begin{equation}
\begin{aligned}
    &\text{PREP}\ket{0}= \sum_j \sqrt{\frac{c_j}{\lambda}}\ket{j}, \text{ where } \lambda=\sum_j c_j\\
    &\text{SEL} \ket{j} = \ket{j} \otimes H_j
\end{aligned}
\end{equation}

Then we have $\bra{0}\text{PREP}^{\dagger}\cdot\text{SEL}\cdot\text{PREP}\ket{0}=H/\lambda$. We note that if no additional constraint is imposed on $H_j$ beyond unitarity, then extra resources are required to construct a self-inverse $\mathrm{SEL}$ operator (see \cref{appx:Qubit}). A self-inverse $\mathrm{SEL}$ operator is necessary for defining the qubitisation operator, which is given by
\begin{equation}
    Q=(2\ket{0}\bra{0}-I)\cdot\text{PREP}^{\dagger}\cdot\text{SEL}\cdot\text{PREP}
\end{equation}
For $n_p=4$, \cref{eqn:ham2} has 18 terms, which means the PREP oracle only needs to prepare a $\lceil\log_218\rceil=5$-qubit state. The cost for this is negligible compared to that of SEL oracle. We also ignore the cost of performing $2\ket{0}\bra{0}-I$ for the same reason. 

Overall, the operator $Q$ has a measurement depth 45184, Toffoli count 91242 and requires 397 logical qubits in total.

The number of queries to $Q$ required to estimate the energy to within error $\sigma$ is given by $\frac{\pi \lambda}{2\sigma}$ \cite{PhysRevX.8.041015}, where $\lambda$ is the sum of spectral norms of the terms in \cref{eqn:ham2}. The number of queries evaluates to $2.44\cdot10^8$, corresponding to a runtime of several years (see \cref{appx:Qubit} for details). Due to this impractical runtime, we omit further analysis on logical qubit breakdown and layout.

Although we now abandon further analysis of qubitisation and report a complete resource audit only for the Trotterisation method, we stress that this does not imply that the Trotterisation is necessarily superior. Our Trotterisation analysis benefitted greatly from our use of the classically validated choice of a relatively large $\delta t$ and the implied modest number of Trotter steps ($10^5$). Had we relied instead on the theoretical bound for $\delta t$ then the Trotter method would also require orders of magnitude longer runtime. Potentially, there may be equivalent arguments to tighten the runtime for qubitisation: the estimate for the number of queries to the block encoding of the Hamiltonian remains a potentially loose upper bound, but no corresponding simulations are currently available to refine it. This may be a direction for future work. Conversely, the qubitisation approach may ultimately prove more costly. Although conceptually straightforward, the first-quantised grid-based representation may not be the most effective choice for qubitisation in our setting, given the relatively small system size.

\section{Applications} \label{sec:applications}


\subsection{Exchange Coupling}

An immediate application is the calculation of the exchange coupling (or $J$-coupling) of the electrons in a DQD. The $J$-coupling is associated with the exchange interaction between electrons and is defined as 
\begin{equation}
    J = E_1-E_0
\end{equation}
where $E_0$ and $E_1$ are the ground state and first excited state energy respectively. The value of $J$, together with how precisely it can be tuned, closely relate to the speed and fidelity of the physical two-qubit gates for spin-qubit platforms \cite{divincenzo2000universal,PhysRevB.83.121403}. The state-of-the-art classical simulation, based on full configuration interaction and implemented in QTCAD \cite{QTCAD_Nanoacademic_2025}, is capable of simulating 4 electrons and extracting the $J$-coupling in approximately one day using 36 vCPUs and 72 GiB of memory.

Computing the $J$-coupling with our quantum algorithm first requires preparing the ground state and then applying QPE to estimate its energy; the first excited state energy can be obtained in the same manner. Therefore, for e.g., $n_p=8$, this would take approximately 18 hours. 

Previous studies \cite{doi:10.1126/sciadv.abo7484,rm43-nh9b} have argued that the cost of state preparation is unlikely to be a major concern. Here, we provided a complementary argument based on the Toffoli count, drawing on a recently proposed method for state preparation. Ref.~\cite{PRXQuantum.6.020319} introduced an approach that enables approximate preparation of a quantum state in a first-quantized plane wave basis for a wavefunction expressed in a Gaussian basis. The Toffoli count of this method scales as $O(n_p^2N_g^{3/2}\log N_w)$, where $n_p$ is the number of particles, $N_g$ is the number of primitive Gaussian basis functions and $N_w$ is the number of plane waves. In their numerical simulations, they found that preparing the Hartree--Fock state of a water molecule requires approximately $10^8$ Toffoli gates, with $n_p = 10$ and $N_w = 8.6\times10^9$. 

For the case of $n_p = 4$ particles in a DQD system, we have $N_w = 6\times10^4$. The corresponding value of $N_g$ is also smaller than that considered in the water-molecule study, but since their results further suggest that the $N_g^{3/2}$ scaling may be overly pessimistic, so we disregard the scaling with $N_g$ in the following discussion. Even then, the corresponding Toffoli cost of state preparation using the same method should be reduced by at least an order of magnitude, yielding a count of $\lesssim 10^7$ Toffoli gates. By comparison, QPE for $n_p = 4$ requires $\sim3\times10^9$ Toffoli gates, indicating that the cost of state preparation is negligible relative to the overall QPE cost.

\subsection{Qubit Shuttling}

Qubit shuttling plays a central role in semiconductor spin-qubit platforms. In these systems, qubits are often spaced further apart to limit crosstalk and to make room for classical control circuitry. As a result, transporting qubits becomes essential for bringing them into close proximity when operations such as entangling gates need to be performed. Classical simulations of single-qubit shuttling have been widely investigated \cite{PhysRevB.102.125406,jeon2025robustness,turner2025modelling}, and their predictions -- including the novel phenomenon that shuttled spins have longer phase lifetimes than static spins -- have been borne out experimentally where speeds and fidelities suitable for fault tolerant QEC have been seen~\cite{LievenShuttle}. However, the in-principle speeds and densities that can be achieved within the device physics limitation are even higher, and advanced ideas such as gating-while-shuttling have been mooted~\cite{JnaneGlobal}. Extending modelling techniques to such scenarios remains challenging and is an opportunity for quantum computation. Indeed, when finer spatial discretization is required to model potential irregularities, such as those induced by charge defects, the memory scaling of quantum algorithms becomes particularly favourable: doubling the grid resolution along a single coordinate doubles the memory requirement of classical methods, whereas quantum grid-based methods require only one additional qubit per particle.

The quantum simulation of multi-qubit shuttling corresponds to evolving a time-dependent extension of the Hamiltonian in \cref{eqn:ham}, which can be addressed within the same Trotter-based simulation framework. In particular, the generalisation of second-order Trotterisation to time-dependent cases is straightforward \cite{Ikeda2023minimum}, whereas applying qubitisation in this context typically requires additional technical effort \cite{Mizuta2023optimalhamiltonian}. The overall cost of simulating the time-dependent Hamiltonian is therefore expected to be comparable to that of the time-independent case.

\section{Conclusion} \label{sec:conclusion}

In this work, we develop a framework for the quantum simulation of multi-particle systems in a quantum dot structures using a grid-based representation. We investigated approaches based on both Trotter-based phase estimation and qubitisation, providing detailed resource estimates for each. Although the theoretical runtime bounds for both methods are impractically large, classical simulations of smaller systems indicate that the Trotterisation approach can, in practice, be executed with orders of magnitude fewer Trotter steps than suggested by worst-case analysis. Based on this observation, the ground-state energy of a four-particle system can be obtained within 22 hours on an early fault tolerant device with 226k noisy qubits using the Trotterisation method. These algorithms offer practical tools for probing exchange coupling and for modelling multi-qubit shuttling dynamics.

These resource estimates were obtained using a standard surface code approach, including the option to store logical qubits in an efficient `yoke' following ref.~\cite{gidney2025yoked}. The resource costs are encouragingly low, but it may be possible to reduce them still further. While finalizing this manuscript, ref.~\cite{low2026denserplanarsurfacecode} appeared which introduced a new surface code approach achieving significantly lower spacetime overheads in, e.g., FeMoco modelling. 
By incorporating these new techniques, together with a redesign of our algorithm and the adoption of a minimum-space layout, the resource cost for borderline classically-intractable quantum dot modelling may fall  below 100k physical qubits without a dramatic increase in runtime.

For future research, it is obviously interesting to explore the adoption of methods from ref.~\cite{low2026denserplanarsurfacecode}. Beyond that, it would be valuable to map out the full layout of our algorithms to verify, for example, that a sufficient number of qubits has been allocated for routing. Another worthwhile direction is to perform classical simulations of the qubitisation method to obtain empirical estimates of the required number of walk-operator queries, which may ultimately yield a spacetime volume smaller than that of the Trotterisation approach. Furthermore, recent advances in using randomised time evolution, and efficient randomised continuous rotation angle compilation techniques \cite{PhysRevLett.132.130602, Kiumi2025,PRXQuantum.5.040352,bothe2026efficientcliffordtsynthesissmallangle}
should be applicable to the resource estimation framework considered here and may lead to substantially lower resource requirements. Finally, our methods can, in principle, be extended to systems beyond DQD. However, encoding an arbitrary three-dimensional background potential \(U\) via a phase oracle is generally costly, and no direct analogue of envelope function theory may exist to simplify this step, potentially leading to an order-of-magnitude increase in runtime. Nonetheless, it is plausible that certain potentials exhibit symmetries that render them approximately separable, for example \(U \simeq U_{xy} + U_z\), thereby effectively reducing the dimensionality of the potential.  Given the potential for further reductions in resource requirements, we conclude that advanced generations of semiconductor-based quantum technology -- including quantum computers -- may be designed with the aid of their own early predecessors.

\section*{Acknowledgement}

The authors thank Po-Wei Huang, Hamza Jnane, Richard Meister and James Williams for fruitful and inspiring discussions.
The authors acknowledge support from the EPSRC QCS Hub grant (agreement No. EP/T001062/1), EPSRC’s Robust and Reliable Quantum Computing (RoaRQ) project (EP/W032635/1), and the SEEQA project (EP/Y004655/1). B.K. thanks UKRI for the Future Leaders Fellowship
project titled Theory to Enable Practical Quantum Advantage (MR/Y015843/1).

\appendix
\section{Variable names and functions}

\begin{itemize}
    \item $n_p$: The number of particles in a double quantum dot system.
    \item $n_q$: The number of qubits per coordinate used to represent a particle in real space.
    \item $n_{QROM}$: The number of qubits used to represent a parameter in the linear piecewise approximated function implemented by QROM phase oracle.
    \item $S_1$: The number of pieces required for the piecewise approximation of the double quantum dot potential.
    \item $S_2$: The number of pieces required for the piecewise approximation of the Coulomb potential.
    \item $l_i$: The flags of an oracle implementing a function with $S_i$ pieces are $l_i$-controlled Toffoli gates, where $l_i=\log_2(S_i)$.
    \item $R_T$: The number of T states required to synthesise a single qubit rotation gate.
    \item $d$: The distance of a quantum error correcting code.
    \item $D_\text{ARI}(n)$: Measurement depth of an $n$-qubit arithmetic operation.
    \item $Q_\text{ARI}(n)$: Extra ancilla required by an $n$-qubit arithmetic operation, which excludes the qubits in the input registers. 
    \item $M_\text{ARI}(n)$: Toffoli count of an $n$-qubit arithmetic operation.
\end{itemize}

We use the in-place adder in ref.~\cite{takahashi2009quantum} with cost metrics:
\begin{itemize}
    \item $D_\text{ADD}(n)=2n-1$
    \item $Q_\text{ADD}(n)=1$
    \item $M_\text{ADD}(n)=2n-1$
\end{itemize}

We use the in-place multiplier in ref.~\cite{li2022circuit} with cost metrics:
\begin{itemize}
    \item $D_\text{MUL}(n)=2n^2+2n+2$
    \item $Q_\text{MUL}(n)=n+1$
    \item $M_\text{MUL}(n)=4n^2+7/2n$
\end{itemize}

We use a squaring based on the in-place multiplier, with cost metrics:
\begin{itemize}
    \item $D_\text{SQR}(n)=2n^2+2n+2$
    \item $Q_\text{SQR}(n)=2n+1$
    \item $M_\text{SQR}(n)=4n^2+7/2n$
\end{itemize}

\section{Resource estimation for phase estimation with Trotterisation}
\label{appx:Trotter}
In this appendix, we provide details of the implementation and resource estimation for the Trotter-based QPE. The code used to compute the resource costs is available upon request. As a representative example, the costs are evaluated explicitly for the case $n_p = 4$ and $n_q = 8$. When selecting among different implementation options (e.g., adder constructions), we are guided by the need to balance qubit efficiency against runtime performance.

\subsection{Stage A: Kinetic}
We first describe the action required on one of the coordinates of a single qubit register and calculate the cost for $2n_p$ registers ($n_p$ particles each with 2 coordinates) at the end.

As explained in the \cref{subsec:trotter}, the strategy is to apply the phase through rotations on bits. For an $n_q$ qubit register, this requires $n_q$ $R_Z$ gates and $n_q(n_q-1)/2$ $CR_z$ gates. Each $CR_z$ gate is composed of 1 relative phase Toffoli, 1 $R_Z$ gates and 1 measurement, requiring 1 ancilla qubits (\cite{nam2020approximate}, fig 2). The mixed fallback method (uses 1 ancilla) \cite{Kliuchnikov:2022rqw} for Clifford+T synthesis of $R_Z$ to precision $\epsilon$ has an average T-count (also T-depth) 
\begin{equation} 
R_T=0.53\log_2(1/\epsilon)+4.86 
\end{equation} 
and $R_T/2$ would give the equivalent Toffoli count. We target an approximation error of $\epsilon = 10^{-4}$, which corresponds to a synthesis depth of approximately 12 for each rotation. For an $n_q$-qubit register, the depth of all the rotations is 
\begin{equation} 
R_T+(R_T+2)(n_q-1)
\end{equation}

The number of ancilla qubits required is $n_q$ (1 qubit for each $R_Z$ in parallel, 2 qubits for each $CR_Z$ in parallel), the ancilla are clean after each rotation.

The Toffoli count for the rotations are from $n_q$ $R_Z$ gates and $n_q(n_q-1)/2$ $CR_Z$ gates, plus each $CR_Z$ gate requires 1 Toffoli, which is given by
\begin{equation}
    n_q\frac{R_T}{2} + \frac{n_q(n_q-1)}{2}(1+\frac{R_T}{2})
\end{equation}

The QFT can be implemented by preparing a reusable phase gradient state and apply additions. The preparation cost of a phase gradient state using the method of ref.~\cite{gidney2025factor} is negligible compared to the other costs and is therefore neglected. A QFT on $n_q$ qubits can be implemented by performing 2-bit addition, 3-bit addition, ..., up to $n_q$-bit in-place addition onto the phase gradient state. Using Gidney's logical-AND adder \cite{Gidney2018halvingcostof}, the total depth is 
\begin{equation} 
D_{\text{QFT}}(n_q)=\sum_{f=2}^{n_q} 4f-4= 2n_q^2-2n_q 
\end{equation}

The number of ancilla required for a $n_q$-bit phase gradient addition is $Q_{\text{QFT}}(n_q)=2n_q-2$ ($n_q-1$ from the `controls' \cite{nam2020approximate}, fig 3 and $n_q-1$ from Gidney's adder).

An $n_q$-bit phase gradient addition requires $n_q-1$ Toffoli for the `controls' and $4n_q-4$ T states for the adder, the total Toffoli count for QFT is thus 
\begin{equation}
    M_{\text{QFT}}(n_q)=\sum_{f=2}^{n_q} \frac {1} {2}(4f-4)+f-1 = \frac {3} {2}n_q^2-\frac {3} {2}n_q
\end{equation}

All analysis above is for a single particle register in one dimension, so for the extra ancilla qubits (in this case QFT uses more ancilla) and Toffoli count, we need to multiply by $2n_p$ to get the total resource requirement for $n_p$ particles each with 2 coordinates. 

Overall, for the case $n_p = 4$ and $n_q = 8$, stage A has depth 333 layers of lattice surgery, requiring extra 112 ancilla qubits and 3281 Toffoli. Note that these estimates do not include the overhead associated with adding the control operations required for phase estimation and therefore do not coincide with the corresponding values reported in \cref{tab:breakdown}.

\subsection{Stage B: DQD potential}
We first describe the action required on a single qubit register and then compute the cost for $n_p$ registers at the end.

The QROM phase oracle for this task is fully serialized to save qubits. The oracle needs to compute and uncompute $S_1$ flags in series, each flag computation/uncomputation has depth $2\lceil\log_2l_1\rceil$ (where $l_1=\log_2S_1$), requiring $2l_1-2$ Toffolis and $l_1-1$ ancilla (clean after each flag computation/uncomputation), while loading the parameters $a_i,b_i,c_i$ requiring only Clifford gates. After all the flags are computed, the parameters are now loaded onto the parameter registers, each of size $n_{QROM}=n_q$. Then the oracle needs to arithmetically compute $U_{xy}^{(i)} = a_ix+b_iy+c_i$, requiring 2 $n_q$-bit multiplication in parallel, followed by 1 $n_{QROM}$-bit addition and 1 $n_q+n_{QROM}$-bit addition. Then an $n_q+n_{QROM}+2$-bit addition on the phase gradient state is required for proper phase implementation. We then run the circuits for calculating $U_{xy}^{(i)}$ in reverse for uncomputation, followed by flag uncomputations.

Using $D$ as the depth function of the arithmetic operations, the measurement depth of this phase oracle is
\begin{equation} 
\begin{aligned}
&4S_1\lceil\log_2l_1\rceil+D_{\text{ADD}}(n_q+n_{QROM}+2)\\
&{+}2[D_{\text{MUL}}(n_q){+}D_{\text{ADD}}(n_{QROM}){+}D_{\text{ADD}}(n_q{+}n_{QROM})]
\end{aligned}
 \end{equation}

Using $Q$ as the extra ancilla count function of the arithmetic operations, the extra ancilla count of this phase oracle is 
\begin{equation} 
\begin{aligned}
&3n_{QROM}+2Q_{\text{MUL}}(n_q)+Q_{\text{ADD}}(n_{QROM})\\
&+Q_{\text{ADD}}(n_q+n_{QROM})+Q_{\text{ADD}}(n_q+n_{QROM}+2)
\end{aligned}
 \end{equation}

Notably, this stage exhibits the highest ancilla requirement throughout the algorithm.

Using $M$ as the Toffoli count function of the arithmetic operations, the Toffoli count of this phase oracle is
\begin{equation} 
\begin{aligned}
&4S_1(l_1-1)+M_{\text{ADD}}(n_q+n_{QROM}+2)\\
&{+}2[2M_{\text{MUL}}(n_q){+}M_{\text{ADD}}(n_{QROM}){+}M_{\text{ADD}}(n_q{+}n_{QROM})]
\end{aligned}
 \end{equation}

\begin{figure*} 
    \centering
    \includegraphics[width=0.9\linewidth]{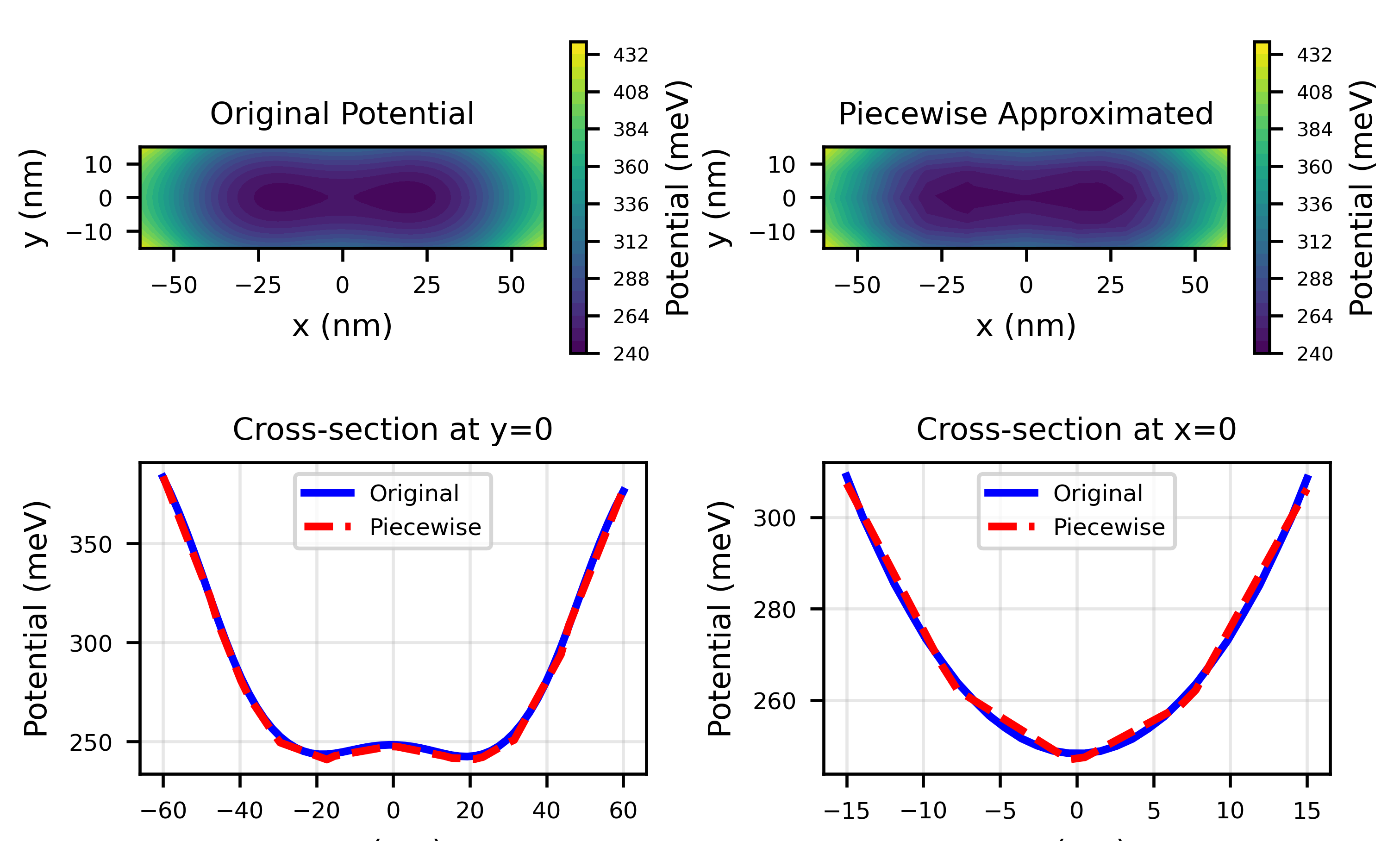}
    \caption{Top: Contour plots of the original potential and its piecewise approximation. Bottom: Cross-sections of the potential and corresponding approximation taken at $y=0$ and $x=0$.}
    \label{fig:potential}
\end{figure*}

For the purpose of estimating the $J$-coupling, we find that \(S_1 = 42\) suffices to achieve the desired accuracy. The exact potential and its piecewise approximation with \(S_1 = 42\) are shown in \cref{fig:potential}.

Overall, for the case $n_p = 4$ and $n_q = 8$, stage B has depth 1028 layers of lattice surgery, requiring extra 180 ancilla qubits and 8412 Toffoli. Note that these estimates do not include the overhead associated with adding the control operations required for phase estimation.

\subsection{Stage C: Coulomb potential}
This stage involing the interactions between particles and thus the operation between registers. We first calculate the squared distance $r^2$ between two particles and then use phase oracle to put up the Coulomb potential energy by approximating a function $f(x)\propto 1/\sqrt{x}$. We need to repeat this process for each particle pair. For $n_p$ particles (assuming even $n_p$) we have $n_p(n_p-1)/2$ pairs and we can do $n_p/2$ in parallel each time, so $n_p-1$ layers of operation is required. 

The arithmetic for calculating the squared distance ($(x_i-x_j)^2+(y_i-y_j)^2$) between two particles is composed of 2 $n_q$-bit subtraction (equivalent to addition) in parallel, 2 $n_q+1$-bit squaring in parallel and 1 $2n_q+2$-bit addition, with the final result storing in a $2n_q+3$-qubit register. The uncomputation runs these circuits in reverse.

The measurement depth for this calculation (including uncomputation) is
\begin{equation} \label{eqn:D_ari}
2(D_{\text{ADD}}(n_q)+D_{\text{SQR}}(n_q+1)+D_{\text{ADD}}(2n_q+2)) \end{equation}


The extra ancilla required for this calculation (including uncomputation) is 
\begin{equation} \label{eqn:Q_ari}
2Q_{\text{ADD}}(n_q)+2Q_{\text{SQR}}(n_q+1)+Q_{\text{ADD}}(2n_q+2) \end{equation}


The Toffoli count for this calculation (including uncomputation) is
\begin{equation} \label{eqn:M_ari}
2(2M_{\text{ADD}}(n_q)+2M_{\text{SQR}}(n_q+1)+M_{\text{ADD}}(2n_q+2)) \end{equation}


Now the register is ready for the phase oracle, for this task since the number of pieces $S_2=8$ is small, the fully serialized ROT1 phase oracle is more advantageous. It operates as follows: computing the flag for piece 1, applying rotations, flag uncomputation, then repeat these for piece 2, piece 3, etc. For flag computation, we keep the ancilla that store the intermediate results, so that the depth and Toffoli count for both computation and uncomputation is halved, at the expense of $l_2-1$ extra ancilla for a $l_2$-controlled flag.

The depth for this oracle is
\begin{equation} \label{eqn:D_ora}
2S_2\lceil\log_2l_2\rceil+(S_2+1)R_T \end{equation}

The extra ancilla count for this oracle is
\begin{equation} \label{eqn:Q_ora}
2n_q+3+l_2 \end{equation}


The Toffoli count for this oracle is
\begin{equation} \label{eqn:M_ora}
2S_2(l_2-1)+(S_2+1)(2n_q+4)R_T/2 \end{equation}


Then we add things up and account for $n_p-1$ layers and $n_p/2$ pairs each layer.

Overall, for the case $n_p = 4$ and $n_q = 8$, stage C has depth 1809 layers of lattice surgery, requiring extra 126 ancilla qubits and 15931 Toffoli. Note that these estimates do not include the overhead associated with adding the control operations required for phase estimation.

\subsection{Overall logical and physical cost}

The above analysis is for one Trotter step in a 2nd order product formula. 
However, phase estimation requires the controlled version of the time evolution operator. The methods for making each Trotter step controlled are:

\begin{itemize}
    \item STAGE A: all the rotations (controlled rotations) become controlled rotations (controlled controlled rotations) and then the QFT and its inverse cancelled itself. Making a rotation into controlled rotation requires 1 extra Toffoli, 1 extra ancilla and 2 extra measurement depth, where we assumed measurement based uncomputation. Similary, making a controlled rotation into controlled controlled rotation requires 1 extra Toffoli, 1 extra ancilla and 2 extra measurement depth.
    \item STAGE B: just after the computation of $U_{xy}$ and before implementing phase by addition, add CSWAP gates controlled on the QPE control qubit and targeted on the register storing $U_{xy}$ and an ancilla register in $\ket{0}$ state, so that if the CSWAP gates are on we just add 0 and no phase is assigned, otherwise it just proceed as usual. We also need another group of CSWAP gates for uncomputaion. All $n_p$ registers can share the same $\ket{0}$ state by staggered the phase implementation, which can slightly slow down the computation.
    \item STAGE C: only the rotations in the phase oracle become controlled rotations; the other parts (arithmetic, compute and uncompute) just cancelled. 
\end{itemize}

This results in marginal increase in both qubit count and runtime, and the results are reported in \cref{tab:breakdown}.

To give physical resource estimates, we assume a 1:1 magic depth to Cliffords depth ratio. Therefore, from \cref{tab:breakdown}, the total depth for one Trotter step should be 6768. We assume the 8T-to-CCZ factory from ref.~\cite{gidney2025factor}, which operates on 12 logical qubits and produces one CCZ state in $\sim5d$ code cycles. Therefore, the average number of factories required to keep up with beat-limited computation is $29064/6768\times5\approx21.5$, corresponding to an average of 257 logical qubits. We also allocate the equal number of qubits to account for routing overhead. Note that the state of the system is stored in  $n_p(2n_q+1)$ data qubits, which is 68 logical qubits for the case $n_p=4$ and $n_q=8$. Furthermore, if we requires $10^{-4}$ precision in the QPE outcome, the number of QPE control qubits required is given by $\lceil \log_2(10^{-4}) \rceil = 14$.

Therefore, the overall logical qubit count is the sum of the contribution from data, ancilla, factory, routing and QPE controls, which is 
\begin{equation} 68+198+257+257+14 = 794 \end{equation}

Assuming physical error rate 0.001, $1\ \mu s$ code cycle time and code distance $d=23$, the QPE algorithm which includes $10^5$ Trotter steps has runtime
\begin{equation} 6768\cdot23\cdot10^{-6}\cdot10^5/3600=4.324 \text{ hours} \end{equation}

From figure 6 in ref.~\cite{gidney2025factor}, we know surface code with code distance $d=23$ has a logical error rate $10^{-14}$ per logical qubit per code cycle. The probability of logical-error-free simulation is thus
\begin{equation} (1-10^{-14})^{794\cdot23\cdot6768\cdot10^5}\sim88.4\% \end{equation}
which justifies out choice of code distance. The expected runtime is thus $4.32/0.884\sim4.89$ hours. Note that this estimate implicitly assumes that the algorithm output can be efficiently verified, for example by comparing it against a classically obtained approximate value, thereby allowing runs affected by logical errors to be identified and discarded. We further neglect the runtime overhead associated with imperfect initial state preparation, which may cause the QPE procedure to return the energy of an excited state rather than that of the target ground state.

The number of physical qubits required per logical qubit is $2(d+1)^2$. Therefore, the total physical qubit count is
\begin{equation} 794\cdot2\cdot(23+1)^2=914688 \end{equation}

Overall, the QPE algorithm requires 794 logical qubits, corresponding to approximately $9.15 \times 10^{5}$ physical qubits at a physical error rate of 0.001, with an expected runtime of about 4.89 hours.

To reduce the physical qubit count, we can slow down the computation by a factor of 2 by serializing the computation: instead of addressing all $n_p$ particle registers in parallel, we address $n_p/2$ registers each time. Consequently, the numbers of logical qubits allocated for ancilla, factory and routing are all halved, which are 99, 128 and 128, respectively. In this configuration, the QPE algorithm has a runtime of 8.65 hours and requires 437 logical qubits, corresponding to $5.03 \times 10^{5}$ physical qubits at a physical error rate of 0.001 (for $d = 23$). The probability of a logical-error-free simulation is 87.3\%, and thus the expected runtime is 9.91 hours. 

For the small footprint scheme, we assume the availability of only 5 magic state factories and employ yoked surface codes to store data qubits. Following figure.~6 of ref.~\cite{gidney2025factor}, 430 physical qubits per logical qubit are sufficient to achieve the same logical error rate as a standard surface code patch with distance $d=23$ (and similarly for $d=25$), corresponding to an approximately threefold reduction in physical qubit count. We further assume that the number of physical qubits allocated to routing is equal to that allocated to data qubit storage, since the reduced parallelization requirements of this scheme substantially alleviate routing overhead. Resource estimates for all remaining components are obtained using the same methodology as described above. Under these assumptions, the total number of physical qubits required is reduced to $2.26\times 10^{5}$.

\subsection{Trotter error bound} \label{subappx:error}

For Hamiltonian in the form $H=A+B$, ref.~\cite{PhysRevX.11.011020} derives a tight error bound for the second-order Suzuki-Trotter formula $\mathscr{S}_2(t)=e^{-i(t/2)A}e^{-itB}e^{-i(t/2)A}$, 
\begin{equation}
\big\|\mathscr{S}_2(t)-e^{-iHt}\big\| \leq 
\frac{t^3}{12} \, \big\| [B,[B,A]] \big\|
+ \frac{t^3}{24} \, \big\| [A,[A,B]] \big\|
\end{equation}
where $\| .\|$ denotes the spectral norm and $[\ ,\ ]$ denotes the commutator. By expanding the commutators and applying $\|A+B\|\leq \|A\|+\|B\|$ and $\|AB\|\leq \|A\|\|B\|$, we can derive a bound that is more straightforward to compute,
\begin{equation} \big\|\mathscr{S}_2(t)-e^{-iHt}\big\| \leq
\frac{t^3}{3} \|B\|^2\|A\|+\frac{t^3}{6}\|A\|^2\|B\| \end{equation}

Using Theorem 1 in supplementary information of ref.~\cite{doi:10.1073/pnas.1619152114}, we can convert the error in time evolution operator to error in energy. Informally, the theorem states that if $\big\|\mathscr{S}_2(t)-e^{-iHt}\big\| \leq \Delta E(t)t$ for $\Delta E(t)$ a non-decreasing continuous function of $t$, then the error in ground-state energies due to Trotterisation is at most $\Delta E(t)$. Thus, 
\begin{equation} \Delta E(t)= \frac{t^2}{6\hbar^2} (2\|B\|^2\|A\|+\|A\|^2\|B\|) \end{equation}
where $\hbar$ is reintroduced to restore dimensions.

Denote the number of Trotter steps as $N$, total simulation time as $T$ and time step as $\delta t = T/N$. The requirement on $N$ is then
\begin{equation} \label{eqn:N_bound}
N^2=\frac{T^2}{6\hbar^2\Delta E} (2\|B\|^2\|A\|+\|A\|^2\|B\|) 
\end{equation}

Recall the form of Hamiltonian in \cref{eqn:ham} and \cref{eqn:ham2}, in our case, we have $A=K=\sum_{i=1}^{n_p}K_x^{(i)}+K_y^{(i)}$ and $B=U+V=\sum_{i=1}^{n_p}U^{(i)}+\sum_{i,j=1;i\neq j}^{n_p} V^{(ij)}$. In grid-based method, the momentum $k_i$ takes integer values between $-2^{n_q-1}$ to $2^{n_q-1}-1$, therefore
\begin{equation} \label{eqn:k_norm}
\sum_{i=1}^{n_p} \big\| K_x^{(i)}\big\| + \big\| K_y^{(i)}\big\|=n_p\cdot \frac{4\pi^2\hbar^2}{mL^2}(2^{n_q-1})^2 
\end{equation}

Since we only capture momentum up to a finite value, the basis functions in real space are ``smeared'' Dirac delta, in the shape of $\frac{\sin x}{x}$. Therefore, each `grid point' has a finite area, which is $(L/2^{n_q})^2$. Thus
\begin{equation} \label{eqn:v_norm}
\sum_{i,j=1;i\neq j}^{n_p}\| V^{(ij)}\| =\frac{n_p(n_p-1)}{2} \frac{e^2}{4\pi \epsilon_0\epsilon_r} \frac{2^{n_q}}{\sqrt{2}L} 
\end{equation}

For $n_p=4$ and $n_q=8$, \cref{eqn:k_norm} evaluates to $13690.8\text{ meV}$, and \cref{eqn:v_norm} evaluates to $1095.2\text{ meV}$. While $\sum_{i=1}^{n_p} \| U^{(i)}\|$ is computed numerically as the difference between maximum and minimum potential, yielding $762.4 \text{ meV}$. 

Evaluating these norms and combining them with \cref{eqn:N_bound} gives a bound on the number of Trotter steps required. For $T=10$ ns and $\Delta E=10^{-4}$ meV, $N\leq 4.13\cdot 10^{11}$ steps. However, it is well known that the analytical bounds for the Suzuki–Trotter formula often overestimate the actual error, sometimes yielding gate count estimates several orders of magnitude greater than those empirically required to reach a specified accuracy \cite{doi:10.1073/pnas.1801723115,PhysRevA.91.022311}. In our case, from classical simulations of $n_p=2$ instantiation, we find $N\sim 10^5$ Trotter steps for QPE should give a good estimation of the ground state energy.

\section{Resource estimation for Qubitisation method}
\label{appx:Qubit}

In this appendix, we discuss the details of block encoding the first quantised Hamiltonian in grid-based representation, as given in \cref{eqn:ham}. We note that the method and cost of simulating first quantised molecular Hamiltonian in a grid-based representation were previously discussed in Appendix K in ref.~\cite{PRXQuantum.2.040332}.

We first block encode the normalized $K_x^{(i)}$ and $K_y^{(i)}$, denoted as $\tilde{K}_x^{(i)}$ and $\tilde{K}_y^{(i)}$. The cost of this part is composed of QFT (and its inverse), a $n_q$ bit squaring (and uncomputation) and $n_q$ $CR_Y$ gates. The cost of a $CR_Y$ gate is similar to that of a $CR_Z$ as $R_Y(\theta)=SHR_Z(\theta)HS^\dagger$. The measurement depth is thus
\begin{equation}
    2D_{\text{QFT}}(n_q)+2D_{\text{SQR}}(n_q)+n_q(R_T+2)
\end{equation}

The number of extra ancilla required is
\begin{equation}
     Q_{\text{SQR}}(n_q)+2
\end{equation}

The Toffoli count is 
\begin{equation}
    2M_{\text{QFT}}(n_q)+2M_{\text{SQR}}(n_q)+ n_q(R_T/2+1)
\end{equation}

The DQD potential is block encoded using the amplitude oracle. Empirically, the widest block encoding is often the Coulomb potential and for $n_q=8$, we numerically verified that the amplitude oracle can be 2-fold parallelised without incur extra ancilla count. The depth of this block encoding is thus 
\begin{equation}
    \frac{1}{2}((S_1+1)R_T + 4S_1\lceil\log_2l_1\rceil)
\end{equation}

The extra ancilla required is
\begin{equation}
    2\max(2n_q +2, l_1-1) + 2n_q+1
\end{equation}
where the $2n_q+1$ term arises from fanning out the data qubits for parallelisation.

The Toffoli count is
\begin{equation}
    (S_1+1)(2n_q + 1) \frac{R_T}{2} + 2S_1(l_1-1)
\end{equation}

The block encoding of the Coulomb potential is constructed analogously to that used for Trotterisation: $r^2$ is first evaluated, followed by application of the amplitude oracle. The resource costs are identical to those reported in \cref{eqn:D_ari,eqn:Q_ari,eqn:M_ari,eqn:D_ora,eqn:Q_ora,eqn:M_ora}.


Once the normalized components \(\tilde{K}_x^{(i)}\), \(\tilde{K}_y^{(i)}\), \(\tilde{U}^{(i)}\), and \(\tilde{V}^{(ij)}\) of the Hamiltonian are encoded separately, we combine them using LCU (PREP and SEL). For an LCU comprising \(u\) terms, \(\log_2 u\) ancilla control qubits are required. In our case, the LCU contains \(n_p^2/2 + 5n_p/2\) terms; for \(n_p = 4\), this corresponds to only 18 terms. Consequently, the cost of the PREP operator is negligible compared to that of SEL, and we therefore neglect it in our resource estimates. The overall cost of the quantum walk operator is thus dominated by SEL. The resulting LCU takes the form
\begin{equation} 
\begin{split}
\frac{1}{\lambda}\Biggl[ & \sum_i \| K_x^{(i)}\|
\begin{pmatrix} 
\tilde{K}_x^{(i)}& * \\
* & * 
\end{pmatrix}
+\|K_y^{(i)}\|
\begin{pmatrix} 
\tilde{K}_y^{(i)}& * \\
* & * 
\end{pmatrix}\\ 
&+\| U^{(i)}\|
\begin{pmatrix}
\tilde{U}^{(i)}& * \\
* & * 
\end{pmatrix}
+\sum_{i,j} \| V^{(ij)}\|
\begin{pmatrix}
\tilde{V}^{(ij)}& * \\
* & *  
\end{pmatrix}  \Biggr]
\end{split}
\end{equation}
with $\lambda=\sum_{i=1}^{n_p} \| K_x^{(i)}\|+\| K_y^{(i)}\|+\| U^{(i)}\|+\sum_{i,j}\| V^{(ij)}\|$ and $\| .\|$ denotes the spectral norm. We recall that these terms contributing to \(\lambda\) were addressed separately in \cref{subappx:error}. The resulting calculation yields  $\lambda=1.55\times10^4$ meV.

Since the terms in \(H\) in \cref{eqn:ham2} are not Hermitian in general, Lemma 10 of ref.~\cite{low2019hamiltonian} implies that making the SEL operator self-inverse incurs a cost approximately equal to two queries to the SEL operator itself. Consequently, constructing a self-inverse SEL operator for the quantum walk requires a total of 399 logical qubits, 91114 Toffoli gates, and 44672 layers of lattice surgery.

For the calculation of the J coupling, one requires at least $\sigma=10^{-4}$ accuracy, this unfortunately results in $\frac{\pi \lambda}{2\sigma}=2.44\times10^8$ calls to the block encoding of $H$. The algorithm runs on 399 logical qubits at a code distance of 29, corresponding to $7.16\cdot10^5$ physical qubits. However, both the runtime and expected runtime are estimated to be 87890 hours, exceeding a decade.

\bibliography{bib}

\end{document}